\documentclass[conference,a4paper]{IEEEtran}
\IEEEoverridecommandlockouts
\usepackage{amsmath,amssymb,amsfonts,mathrsfs}
\usepackage{algorithmic,xspace,booktabs}
\usepackage{graphicx,soul}
\usepackage{textcomp}
\usepackage{xcolor}
\usepackage[normalem]{ulem}
\usepackage[utf8]{inputenc} 
\usepackage[T1]{fontenc}
\usepackage{array}
\usepackage{mathtools}
\usepackage{flushend}

\usepackage[scaled=.85]{beramono}
\usepackage[inline, shortlabels]{enumitem}

\newcommand{\ft}[1]			{\left[\kern-0.15em\left[#1\right]\kern-0.15em\right]}
\newcommand{\fe}[1]		{\left[\kern-0.30em\left[#1\right]\kern-0.30em\right]}
\newcommand{\flr}[1]		{\left\lfloor #1 \right\rfloor}
\def\DE					{\stackrel{\rm{def}}{=}}

\def\usf  {\texttt{USF}\xspace}
\def\opamp  {\texttt{OpAmp}\xspace}

\def\enob  {\texttt{ENOB}\xspace}
\def\sinad  {\texttt{SINAD}\xspace}

\def\DR   {\texttt{DR}\xspace}
\def\DRes   {\texttt{DRes}\xspace}
\def\HDR   {\texttt{HDR}\xspace}
\def\LDR   {\texttt{LDR}\xspace}

\def\madc {$\mathscr{M}_\lambda$-\texttt{ADC}\xspace}
\def\imadc {$\mathscr{M}_\lambda$-${\int}$\texttt{ADC}\xspace}

\newcommand{\bpara}[1]{\smallskip \noindent {\bf #1}}

\newcommand\fig[1]				{Fig.~\ref{#1}}
\newcommand{\MO}{\mathscr{M}_\lambda}
\newcommand\secref[1]			{Section \ref{#1}}
\newcommand{\stp}{$\text{ST}^{+}$}
\newcommand{\stn}{$\text{ST}^{-}$}

\def\BibTeX{{\rm B\kern-.05em{\sc i\kern-.025em b}\kern-.08em
    T\kern-.1667em\lower.7ex\hbox{E}\kern-.125emX}}
\begin{document}

\title{Unleashing Dynamic Range and Resolution in \\ Unlimited Sensing Framework via Novel Hardware \\
\thanks{The work of AB is supported by the UK Research and Innovation council's FLF Program ``Sensing Beyond Barriers via Non-Linearities'' (MRC Fellowship award no.~MR/Y003926/1).}
}

\author{ 
Yuliang Zhu
and
Ayush Bhandari \\ 
Dept. of Electrical and Electronic Engg., 
{Imperial College London}, SW7 2AZ, UK. \\ 
{
\texttt{\{yuliang.zhu19,a.bhandari\}@imperial.ac.uk}} 
\ or  \ 
\texttt{ayush@alum.MIT.edu} \\

{\small \color{red} Proc. of 2024 IEEE SENSORS (Conference)}
}

\maketitle

\begin{abstract}
Conventional digitization based on the Shannon-Nyquist method, implemented via analog-to-digital converters (ADCs), faces fundamental limitations. High-dynamic-range (HDR) signals often get clipped or saturated in practice. Given a fixed bit budget, one must choose between minimizing quantization noise or accommodating HDR inputs. The Unlimited Sensing Framework (USF) eliminates saturation by incorporating nonlinear folding in analog hardware, resulting in modulo signals. Quantizing or digitizing modulo signals enables low quantization noise as the modulo representation maps HDR signals into low-dynamic-range (LDR) samples. In the context of USF, the core innovation of this paper is a novel, low-cost, integrator-based efficient modulo ADC hardware implementation that imposes no restrictions on folding rates, enabling significantly HDR capture. The feasibility of this design is demonstrated by hardware experiments showcasing clear advantages across different quantitative performance metrics. These include capturing HDR signals with a 60-fold increase in dynamic range, achieving up to 5 Effective Number of Bits (ENOBs), and improving Signal-to-Noise and Distortion (SINAD) by 30 dB.

\end{abstract}


\section{Introduction}

\begin{figure}[htbp]    \centerline{\includegraphics[width=1\linewidth]{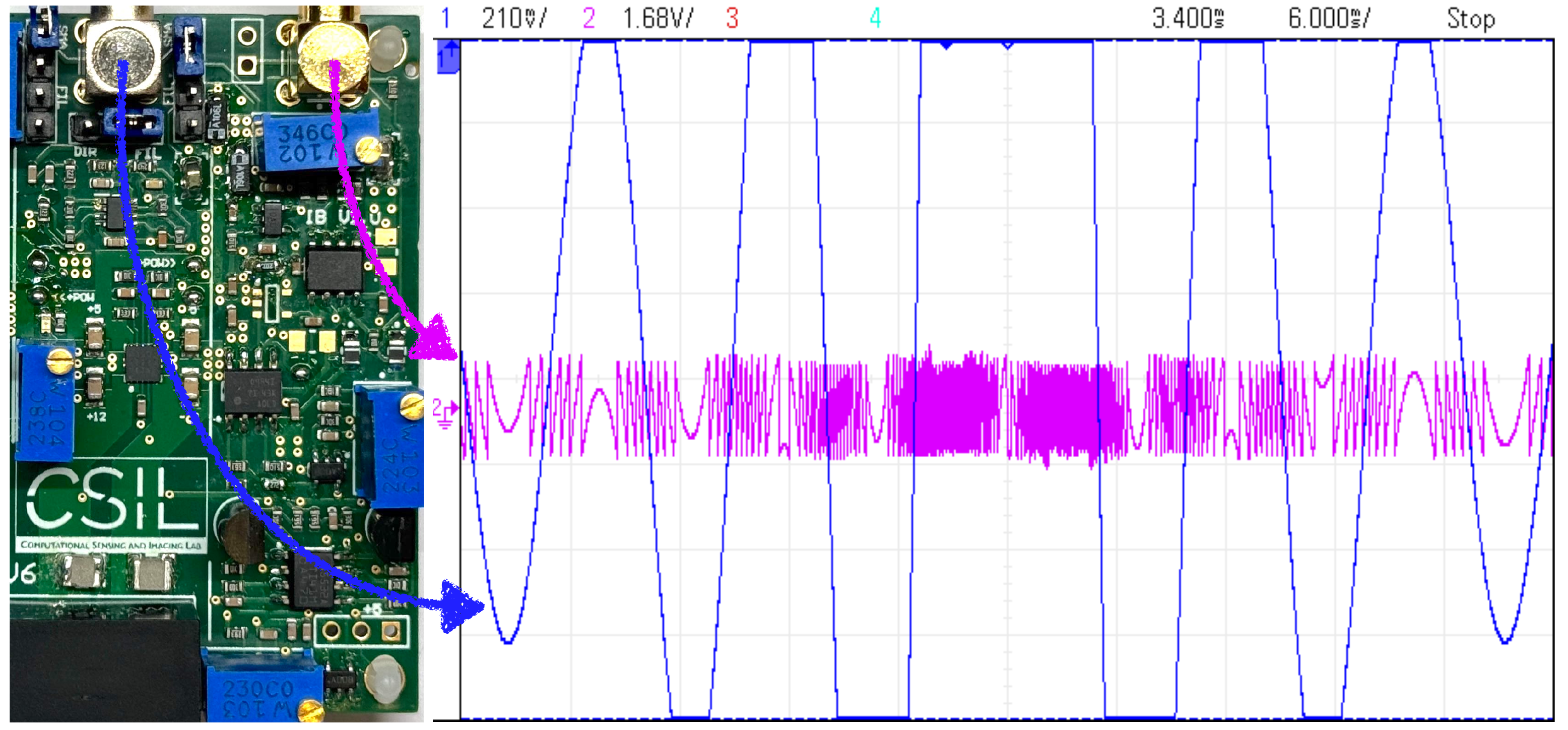}}
    \caption{Hardware prototype for the integrator-based modulo ADC. The initial design is capable of folding a signal as large as \( 20\lambda \), with no theoretical limits on folding times. On the right is a screenshot of the DSOX1204G oscilloscope digitizing a sinc function (Blue) with \( V_{\text{pp}} = 16V \) and the output of the integrator-based modulo ADC (Purple) configured with \( \lambda = 1 \) and \( \alpha = 8 \). The DR gain reached more than 60 times modulo threshold $2\lambda$.}
    \label{fig_pcb}
\end{figure}

Digital acquisition is the key driver of almost all modern technologies and applications. At the heart of this digital acquisition paradigm is the \textit{Shannon-Nyquist} theorem, that defines the minimum sampling rate required to accurately capture band-limited signals. Specifically, it states that signal recovery is achievable if the sampling frequency is at least twice the signal's bandwidth, a threshold known as the Nyquist rate. This principle is the cornerstone of traditional digital acquisition, typically implemented using analog-to-digital converters (ADCs) \cite{Shannon:1948:J}. \uline{Fundamental limitations} in conventional ADCs include, (i) {\bf Saturation} or {\bf Clipping}. If the signal's amplitude exceeds ADC range, it is clipped to a maximum value, resulting in irretrievable information loss, as shown in \fig{fig_pcb}. (ii) {\bf Quantization Noise.} While increasing the ADC's amplitude range can prevent clipping, it leads to poor digital resolution due to coarse quantization. This introduces a fundamental trade-off between Dynamic Range (\DR) and Digital Resolution (\DRes). The Unlimited Sensing Framework (\usf) overcomes these limitations at once \cite{Bhandari:2017:C,Bhandari:2020:Ja,Bhandari:2021:J,Florescu:2022:J,Guo:2023:C,Beckmann:2024:J, Liu:2023:J,Zhu:2024:C,Feuillen:2023:C}.

\bpara{The Unlimited Sensing Framework.} The \usf is a computational sensing paradigm \cite{Bhandari:2022:Book} that leverages a collaboration of both \textit{hardware} \cite{Bhandari:2021:J,Florescu:2022:J} and \textit{algorithms} \cite{Bhandari:2017:C,Ordentlich:2018:J,Bhandari:2021:J,Guo:2023:C}. Before sampling, the analog-domain modulo hardware performs a \uline{folding operation} on the continuous signal via the mapping,
\begin{equation}
    \MO: g \to 2\lambda \left( {\fe{ {\frac{g}{{2\lambda }} + \frac{1}{2} } } - \frac{1}{2} } \right), 
		\ \  \ft{g} \DE g - \flr{g}
    \label{eq1}
    \end{equation}
where $\flr{g}$ is the floor function. Consequently, any high-dynamic-range (\HDR) analog signal is folded back into low-dynamic-range (\LDR) ADC, with a range of $[-\lambda, \lambda]$, thus {\bf eliminating saturation}. Thereon, mathematical algorithms recover the HDR signal from the LDR folded samples using a sampling criterion similar to Shannon-Nyquist rate, independent of $\lambda$ \cite{Bhandari:2017:C,Bhandari:2020:Ja}. Beyond theory,  hardware implementation of modulo ADCs have been presented in literature \cite{Bhandari:2021:J,Florescu:2022:J,Mulleti:2023:J}, demonstrating significant  performance enhancement both in terms of  {\DR} and {\DRes}, across applications spanning 
tomography \cite{Beckmann:2024:J}, 
digital communication \cite{Liu:2023:J,Ordonez:2021:J},
sub-Nyquist sampling \cite{Zhu:2024:C} and
radars \cite{Feuillen:2023:C}.

\bpara{Motivation.} Modulo ADC or \madc presented in existing literature \cite{Bhandari:2021:J,Mulleti:2023:J} typically consist of a counter and a digital-to-analog converter (DAC). These digital components  introduce design challenges as follows.
\begin{enumerate}[leftmargin = *]

    \item \textbf{Bit Limitation.} The counter and DAC usually have a finite number of bits $(B)$, limiting input \DR to $2\lambda (2^B-1)$. This has led to $32\lambda$\cite{Bhandari:2021:J} and $16\lambda$ \cite{Mulleti:2023:J} implementations, respectively. Can we decouple $B$ and $\lambda$? This will enable us to capture truly \HDR signals, unrestricted by $B$.
    
    \item \textbf{Adaptability Issues.} 
    In many cases, adjusting the modulo threshold $\lambda$ is required. This may be for compressing \DR by invoking higher folds or for matching the \madc output voltage to conventional samplers. This requires tuning the reference and offset voltages of the DAC, making it difficult to adaptively vary the gain.

\end{enumerate}

Additionally, the DAC generates impulses, introducing reset noise \cite{Bhandari:2022:C}. The DAC-based \madc is often bulky, power-intensive, and complex, limiting its applicability in various contexts. All of these aspects motivate new designs and approaches for \madc hardware that is at the core of \usf.

\begin{figure}[!t]    \centerline{\includegraphics[width=1\linewidth]{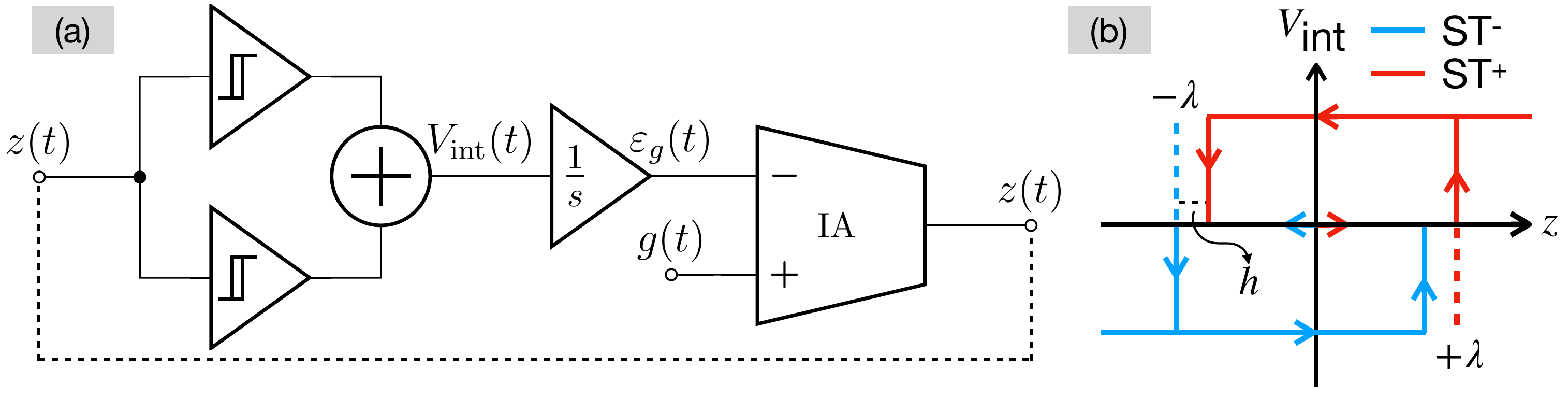}}
    \caption{Architecture of the \imadc. (a) Block diagram of the proposed system. The input is $g(t)$. The output is the modulo signal $z(t)$. (b) Transfer function of the Schmitt trigger pair and adder.}
    \label{fig_sch}
\end{figure}

\bpara{Contributions.} In this paper, we introduce a novel, simple, and cost-effective integrator-based modulo ADC or \imadc, shown in \fig{fig_sch}. The \imadc overcomes the above bottlenecks, offering significant improvements over current art:

\setlength{\fboxsep}{2pt}
\setlength{\fboxrule}{0.5pt}

\begin{enumerate}[leftmargin=*,itemsep = 2pt,label = {\small$\textrm{C}_\arabic*$})]

\item \textbf{\DR Gain.} We design an integrator-based hardware that can produce theoretically an unlimited number of output levels, enabling an unlimited number of folds. Proof-of-concept experiments have demonstrated up to 60 folds, as shown in \fig{fig_pcb}.

\item \textbf{Adaptive \DR Gain.} This unique architecture provides \textit{on-the-fly} adjustment of \DR gain which can be controlled by $\alpha$, independent of the input signal and $\lambda$. This opens up a possibility for incorporating automatic gain control with in \madc architecture, as shown in \secref{sec:exp}, Experiment 2 and \fig{fig5}; this was previously unexplored.
    
\end{enumerate}

The architecture shown in \fig{fig_sch} is practical due to its intrinsically low power design, simplicity, and ease of reproduction, making USF hardware accessible to entry-level university students and interdisciplinary researchers. Furthermore, the low-pass characteristic of the integrator eliminates impulse noise found in previous systems \cite{Bhandari:2022:C}.

\section{System Architecture and Overview}
\label{sec:SAO}

\bpara{Qualitative Introduction to \imadc.}
The novel \imadc architecture is shown in \fig{fig_sch} (a). This design aims to implement the modulo mapping in \eqref{eq1} in an alternative yet advantageous manner compared to previous approaches \cite{Bhandari:2021:J,Florescu:2022:J,Mulleti:2023:J}. Let the band-limited input signal be denoted by $g(t)$ with its modulo counterpart $z(t) = \MO(g(t))$. The input $g(t)$ admits a unique decomposition: $g(t) = z(t) + \varepsilon_g(t) $ where the residue function $\varepsilon_g(t) = 2\lambda \sum_{{l}\in\mathbb{Z}} e_l \mathnormal{1}_{\mathcal{D}_l}(t)$ \cite{Bhandari:2017:C}. In practice, an instrumentation amplifier (IA) is used as a subtractor to output the modulo signal from the modulo decomposition $z(t) = g(t) - \varepsilon_g(t)$. While $g(t)$ is readily accessible, the residue function $\varepsilon_g(t)$ is obtained as the output of the integrator. An ideal operational amplifier or \opamp-based integrator with input signal $V_{\texttt{int}}(t)$ drives the output,
\begin{equation}
    \underbrace{\varepsilon_g(t)}_{\textsf{Output}} = -\frac{1}{RC}\int^{t}_{0} \underbrace{V_{\texttt{int}}(t)}_{\textsf{Input}} \, dt + \varepsilon_g(0).
\end{equation}
The integrator input $V_{\texttt{int}}(t)$ is a sum of a Schmitt trigger (ST) pair with the transfer function shown in \fig{fig_sch} (b). The folding process is initiated by the pair of STs. For instance, 
\begin{enumerate}[leftmargin = *, label = $\bullet$]
    \item Whenever $|z(t)|>\lambda$, \stp (red) generates a positive voltage while \stn (blue) remains neutral (0V). The sum of the STs forces the integrator output $\varepsilon_g(t)$ to rise steeply, which then drives the modulo output $z(t)$, downwards.

\item After $z(t) < (h - \lambda)$, \stp returns to 0V and the integrator stabilizes to a constant voltage, completing the folding cycle. Note that the hysteresis parameter $h$ is manually tuned \cite{Florescu:2022:J} to prevent oscillations and unnecessary jumps.  
\end{enumerate}

The time elapsed for the folding process changes adaptivly with respect to the gain of the integrator and the IA.

\bpara{Unleashing the Dynamic Range (\DR).}
As before, let us quantify the \DR by $\rho=\frac{\max(g(t))-\min(g(t))}{2\lambda}$ \cite{Bhandari:2021:J}. If $|g(t)|<\lambda$, $z(t)$ will be not different from the $g(t)$ because no folding has been triggered. Previously, $\lambda$ was fixed during the operation, and higher $\rho$ was only achievable by increasing the input amplitude. This approach is sub-optimal as amplifying the input signal necessitates higher voltage supply rails and additional power, specially in experimental test-beds.

The hardware presented in this work utilises the following, 
\begin{equation}
    z(t) = \alpha*\mathscr{M}_{\lambda/\alpha}(g(t))=\mathscr{M}_{\lambda}(\alpha*g(t)).
    \label{eq3}
\end{equation}
We introduce $\alpha$ as the new parameter modulating the ``modulo gain.'' Intuitively, adjusting $\alpha$ is equivalent to scaling the input amplitude by $\alpha$. In electronics, $\alpha$ is the gain of the IA which is easily tuned by its feedback resistor value. This enables the implementation that can be seen as a parallel to automatic gain control (AGC) in conventional hardware \cite{AlegrePerez:2011:Book}. In the \usf context, by optimising $\alpha$, one can gain Effective Number of Bits (\enob), see \fig{fig5} in \secref{sec:exp}, opening up new ways to controlling \DRes.

%

\bpara{HDR Signal Recovery.} Though algorithms are not a focus of our discussion, we use an adaptation of USF recovery algorithm \cite{Bhandari:2017:C,Bhandari:2020:Ja}. Direct application of the USF algorithm can lead to its breakdown due to non-idealities \cite{Bhandari:2021:J}. For instance, electronic components have a finite slew rate, which impedes ideal discrete modulo folding (instantaneous transition from $\lambda$ to $-\lambda$ or vice versa) \cite{Bhandari:2020:Ja}. Consequently, the folding transitions are sloped, leading to invalid samples being taken during this transition period. To mitigate this issue, we use a \textit{reset sample correction algorithm}, that employs a sliding window to identify samples taken during the reset by examining the maximum difference within each window. Subsequently, it adjusts the samples taken during the transition to predicted values using linear regression. This method proves highly effective, provided the transition period is brief.


\begin{figure}[!t]
    \centerline{\includegraphics[width=1\linewidth]{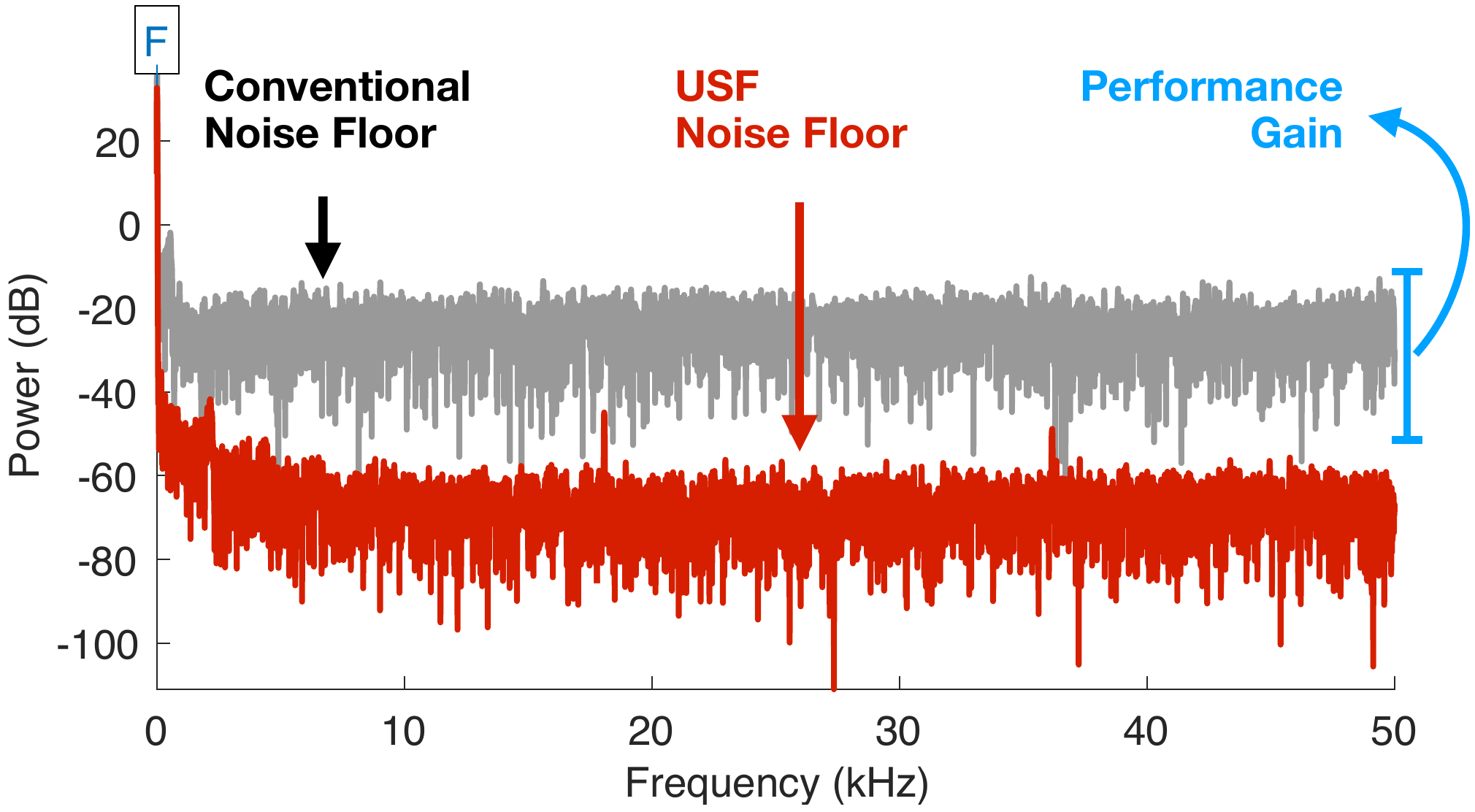}}
    \caption{Signal to Noise and Interference Ratio Comparison between USF and CADC. The resulting 30 dB gain in SINAD is achieved with $\rho = 8$ and $\alpha = 8$, demonstrating an improvement of more than 60-fold. }
    \label{fig_SINAD}
\end{figure}

\begin{figure}[htbp]    \centerline{\includegraphics[width=1\linewidth]{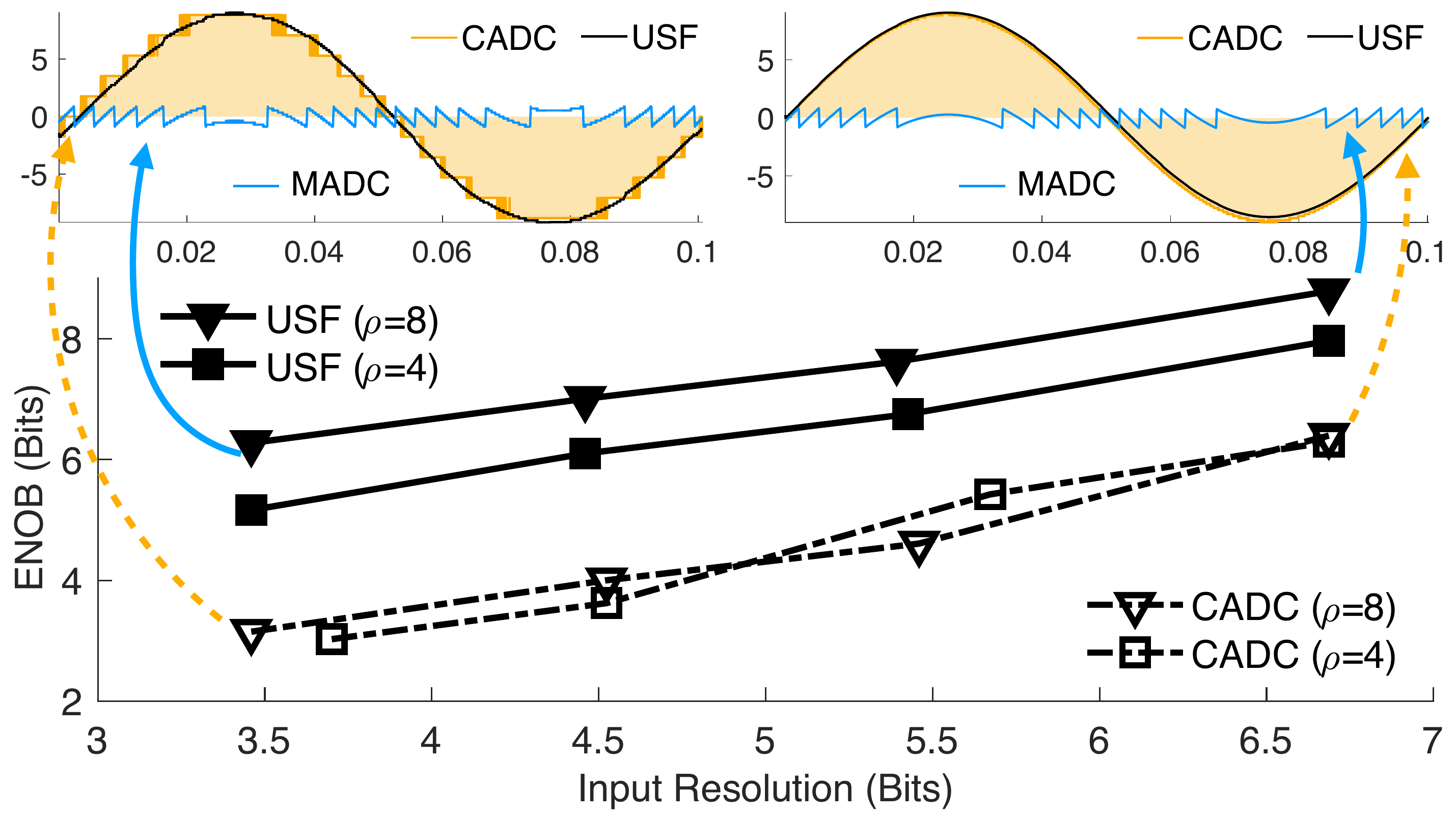}}
    \caption{Performance of \imadc compared with conventional ADC. }
    \label{fig_exp1}
\end{figure}

\section{Experimental Results}
\label{sec:exp}

To investigate the validity of the proposed design, a hardware prototype of the \imadc was designed to implement the \usf pipeline. The electronic circuit together with an oscilloscope screenshot are shown in Fig \ref{fig_pcb}. Overall results, performance metrics and corresponding experimental figures are tabulated in Table~\ref{tab:PM}.

\bpara{Experimental Protocol}
For each experiment, A 10 Hz Sine wave is used as the input signal $g(t) = 2\lambda\rho \sin{(20\pi t)}$. An 8-bit oscilloscope PS3406D is used to simultaneously digitize the input signal (ground truth) and its modulo samples $y[k] \in [-\lambda, \lambda]$ where $\lambda\approx 1$.  The input \HDR signal is reconstructed by the approach in \secref{sec:SAO}. The Signal-to-Noise and Distortion ratio (\sinad) is then computed for both the quantize conventional input and the modulo recovered signal. We use the Effective Number of Bits (\enob) as a direct measure of the output dynamic range.

\bpara{Experiment 1: \DR gain from amplitude range.} 
The goal of the first experiment is to investigate the impact of changing input amplitude on the system's \enob while keeping modulo gain fixed, or $\alpha=1$. Two input amplitudes, $8$V and $16$V, corresponding to $\rho=4$ and $\rho=8$, respectively, are tested with different sampling resolutions in both systems. The results are plotted in \fig{fig_exp1} which demonstrate the clear superiority of \usf approach.

\begin{figure}[!t]
\centerline{\includegraphics[width=1\linewidth]{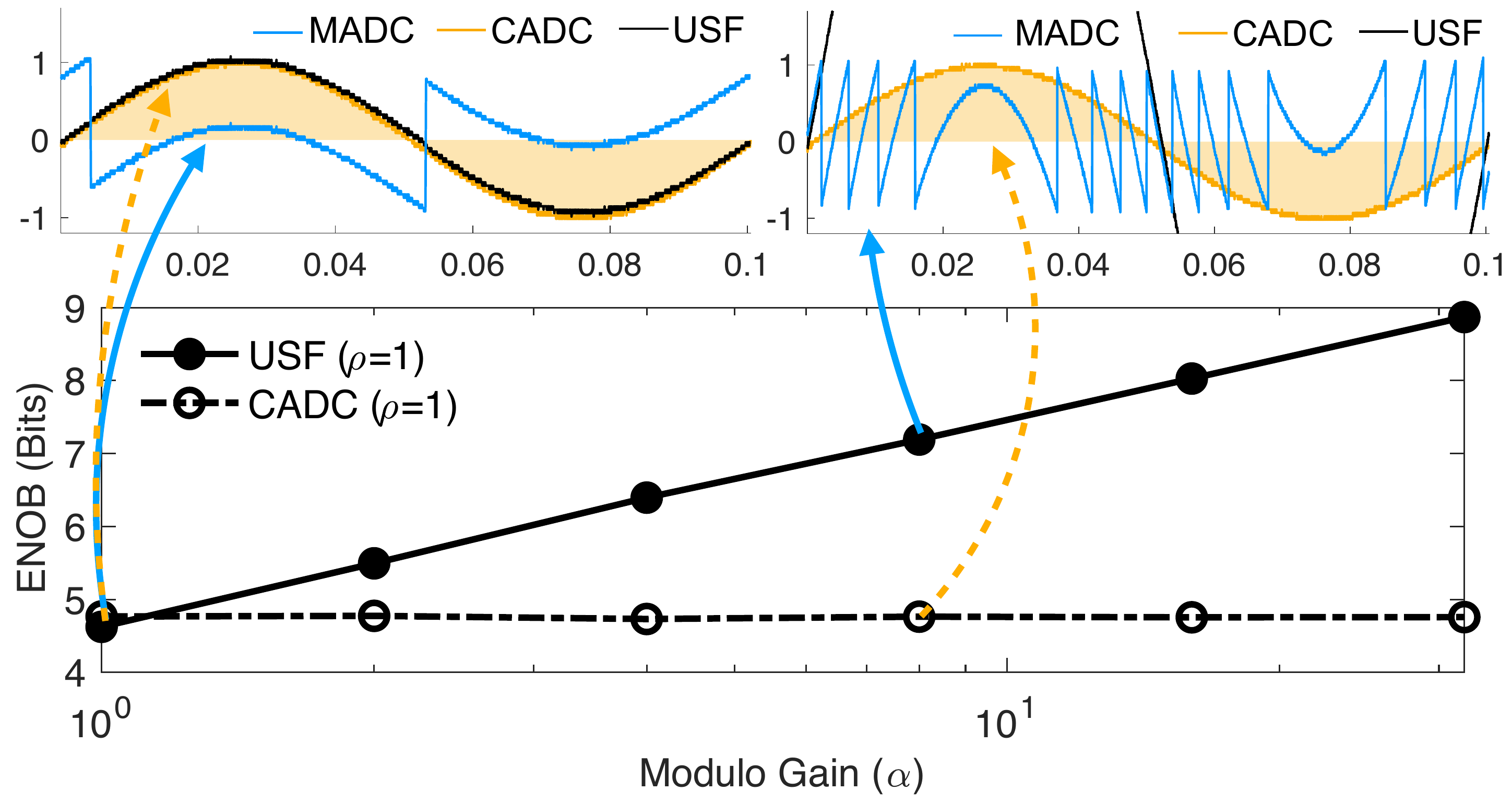}}
    \caption{\enob improvement by varying $\alpha$.}
    \label{fig5}
\end{figure}

\bpara{Experiment 2: \DR gain from modulo gain.} The goal of the second experiment is to show the effect of changing modulo gain $\alpha$ to [1, 2, 4, 8, 16, 32] while setting input amplitude to $\rho = 1$ (2V), demonstrating the dynamic range of the \usf goes beyond the gain from the amplification of the input signal amplitude. The modulo gain $\alpha$ is easily adjusted by changing the gain resistor (potentiometer) of the instrumentation amplifier (AD8421), which is formulated as $\alpha=1+\frac{9.9k \Omega}{R_G}$ \cite{AD8421}. Empirically, our results show that the gain in \enob with varying $\alpha$ can be approximated by $\log_2{\alpha}$. 

\begin{table}[!h]

\centering

\caption{\usf: Performance Metrics}
\label{tab:pm}
\resizebox{0.4\textwidth}{!}{%
\begin{tabular}{@{}cccccc@{}}
\toprule
Figure & $\rho$ & $\alpha$ & $\texttt{SINAD}^{\mathrm{1}}$ & $\texttt{SINAD}^{\mathrm{2}}$ & \sinad Gain \\ \midrule
       &        &          & (dB)                 & (dB)                 & (dB)       \\ \midrule
Fig. 3 & 8      & 8        & 25                   & 56.85                & 31.85      \\
Fig. 4 & 4      & 1        & 25.85                & 37.24                & 11.39      \\
Fig. 4 & 8      & 1        & 25.85                & 43.95                & 18.10      \\
Fig. 5 & 1      & 1        & 30.45                & 29.60                & -0.85      \\
Fig. 5 & 1      & 4        & 30.25                & 40.26                & 10.01      \\
Fig. 5 & 1      & 8        & 30.44                & 45.05                & 14.61      \\
Fig. 5 & 1      & 16       & 30.39                & 50.09                & 19.70      \\
Fig. 5 & 1      & 32       & 30.40                & 55.16                & 24.76      \\ \midrule
\multicolumn{6}{l}{$^{\mathrm{1}}$\sinad of the Conventional ADC.}                     \\
\multicolumn{6}{l}{$^{\mathrm{2}}$\sinad of the \imadc.}                \\ \bottomrule
\end{tabular}%
}
\label{tab:PM}
\end{table}

\section{Conclusion}
The Unlimited Sensing Framework (\usf) is an alternative digital sensing method that can digitize  high-dynamic-range (\HDR) with high-digital-resolution, without clipping or saturating. In this context, modulo ADCs play a crucial role for signal folding. This paper presents a novel integrator-based modulo ADC that significantly enhances the dynamic range capabilities of the \usf. Unlike current modulo ADC architectures limited by digital counters and DACs, our design is not constrained by the number of folds. Our hardware experiments show promising results across different performance metrics. This opens up new possibilities in the context of \usf.

\end{document}